\documentclass[12pt]{article}
\usepackage{amsmath,amssymb,cite,epsfig}
\advance \topmargin by -\headheight
\advance \topmargin by -\headsep     
\evensidemargin \oddsidemargin
\def\Maketitle{{\def\newpage{}\maketitle}}
\makeatletter
\def\Appendix{\appendix
  \def\@seccntformat##1{Appendix~\csname the##1\endcsname.~~}}
\makeatother
\makeatletter
\@addtoreset{equation}{section}

\makeatother
\begin{document}
\title{\textbf{Multipoint correlation functions in \\Liouville field theory and minimal Liouville gravity}\\
\vspace*{32pt}\normalsize{Contribution to the proceedings of the\\ International Workshop on Classical and Quantum Integrable Systems,\\ Dubna, Russia, January 22--25, 2007.}}
\vspace*{12pt}
\author{V.~A.~Fateev$^{1,2}$ and A.~V.~Litvinov$^1$\\[\medskipamount]
$^1$~\parbox[t]{0.85\textwidth}{\normalsize\it\raggedright
Landau Institute for Theoretical Physics,
142432 Chernogolovka, Russia}\\[\medskipamount]
$^2$~\parbox[t]{0.85\textwidth}{\normalsize\it\raggedright
Laboratoire de Physique Th\'eorique et Astroparticules, UMR5207 CNRS-UM2, Universit\'e
Montpellier~II, Pl.~E.~Bataillon, 34095 Montpellier, France}
}
\rightline{PTA-07.15}
\date{}
\Maketitle
\begin{abstract}
We study $n+3$-point correlation functions of exponential fields in Liouville field theory with $n$  degenerate and $3$ arbitrary fields. An analytical expression for these correlation functions is derived in terms of Coulomb integrals. The application of these results to the minimal Liouville gravity is considered.
\end{abstract}
\section{Liouville field theory}\label{Liouville}
Last years since the seminal paper of A.~Polyakov \cite{Polyakov:1981rd} Liouville field theory (LFT) attracts a lot of attention mainly due to it deep relation with the theory of bosonic strings in noncritical dimensions. LFT gives an example of non-rational CFT, which has different physical applications. Some results derived in LFT and minimal Liouville gravity can be compared with the results coming from the matrix models \cite{Dotsenko:1991id,DiFrancesco:1991ud,Zamolodchikov:2005fy,Belavin:2005yj,Belavin:2006ex,Kostov:2005kk,Kostov:2005av}. For the special set of fields (so called degenerate fields) the correlation functions in LFT are simply related to the correlation functions in the minimal models of CFT, which describe critical behavior of many interesting two-dimensional statistical systems. 

Liouville field theory is described by local Lagrangian density
\begin{equation}\label{Lagrangian}
   \mathcal{L}=\frac{1}{4\pi}(\partial_a\varphi)^2+\mu e^{2b\varphi}
\end{equation}
with holomorphic stress-energy tensor
\begin{equation*}
  T(z)=-(\partial_z\varphi)^2+Q\partial_z^2\varphi,
\end{equation*}
which ensures local conformal invariance of the theory with central charge $c_L$, parameterized in terms of coupling constant $b$ as
\begin{equation}
   c_L=1+6Q^2,
\end{equation}
where $Q=b+b^{-1}$.
The scale parameter $\mu$ in Eq \eqref{Lagrangian} is called the cosmological constant. 

Basic objects in this theory are the exponential fields parameterized by a continuous parameter $\alpha$
\begin{equation}\label{primary}
  V_{\alpha}(z,\bar{z})=e^{2\alpha\varphi},
\end{equation}
which are the primary fields of the Virasoro algebra generated by stress-energy tensor $T(z)$ with the conformal dimensions $\Delta_L(\alpha)=\alpha(Q-\alpha)$. Here and later $z$ is a complex coordinate on a plane 
\begin{equation*}
  z=x_1+ix_2.
\end{equation*}
To simplify the notations we write the primary field defined by Eq \eqref{primary} simply as $V_{\alpha}(z)$ and denote
\begin{equation*}
  d^2z=dx_1dx_2.
\end{equation*}
The important property of LFT is that the fields $V_{\alpha}$ and $V_{Q-\alpha}$ have the same conformal dimension and really represent the same conformal field. It means, that they are related by a linear transformation
\begin{equation}\label{Reflection}
   V_{\alpha}=R(\alpha)V_{Q-\alpha},
\end{equation}
with function
\begin{equation*}
  R(\alpha)=\frac{(\pi\mu\gamma(b^2))^{(Q-2\alpha)/b}}{b^2}\frac{\gamma(2b\alpha-b^2)}
  {\gamma(2-2\alpha/b+1/b^2)},
\end{equation*}
which is known as the reflection amplitude. Here and later we use the notation
\begin{equation}
\gamma(x)=\Gamma(x)/\Gamma(1-x).
\end{equation}
In this paper we consider the correlation functions in LFT, which contains three arbitrary and $n$ degenerate fields. Degenerate fields in LFT correspond to the primary fields $V_{\alpha}(z)$ with the value of parameter $\alpha$\footnote{Equation \eqref{degenerate} should be understood modulo transformation \eqref{Reflection}.}
\begin{equation}\label{degenerate}
   \alpha=-\frac{mb}{2}-\frac{nb^{-1}}{2}\;\;\;m,n=0,1,2,\dots.
\end{equation}
To do all formulae of this paper more transparent we consider only  the case $n=0$. The correlation functions with $n\neq0$ have more tedious form and we suppose to consider them in other publication. 

In is well known, that four-point correlation function with degenerate field $V_{-mb/2}(z)$ in LFT satisfy $m+1$ order differential equation in each of the variable $z$ and $\bar{z}$ \cite{Belavin:1984vu}. The explicit form for this function determines the integral representation for the solution to this differential equation. The correlation functions with three arbitrary fields and more than one degenerate fields satisfy already the system of differential equations in partial derivatives and the possibility to write the solution to this system in terms of finite-dimensional integrals is not evident. The solution to the conformal bootstrap problem for these correlation functions in terms of finite dimensional Coulomb integrals is the main result of this paper. In the case when more than three fields are non-degenerate the correlation function contains an infinite number of conformal blocks and there is no reason to 
expect, that it can be written as finite-dimensional integral. In this paper we express the multipoint correlation functions in terms of the integrals over the whole plane. In the region of convergency these integrals define completely correlation functions. Outside this region they should be understood in the sense of analytical continuation. This continuation can be performed for example by rewriting integrals 
over plane in terms of contour integrals, as it was described in \cite{Dotsenko:1984nm,Dotsenko:1984ad} (see also the appendix \ref{4point-integrals}).

The correlation functions in LFT with three arbitrary and $m$ degenerate fields have many different applications. In LFT (or in minimal models of CFT) perturbed by degenerate field they give the possibility to express the perturbative corrections  to the  three-point correlation functions (structure constants of operator product expansion) in terms of finite-dimensional integrals. Even in the minimal models of CFT \cite{Belavin:1984vu}, where for all non-zero correlation functions the screening condition is satisfied, our results give independent integral representation (which in many cases is simpler), where the number of integrations does not depend on three arbitrary fields. These multipoint correlation functions also appear and  play an important role in studying of the conformal Toda field theory. 
We suppose to discuss the application of multipoint correlation functions in LFT to perturbed CFT and conformal Toda field theory in other publication. Here in section \ref{MLG} we briefly consider the application of these functions to the minimal Liouville gravity. 

Multipoint correlation functions in LFT are rather complicated objects, however
as it was noticed in \cite{Goulian:1990qr}, any multipoint correlation function  
$\langle V_{\alpha_1}(z_1)\dots V_{\alpha_m}(z_m)\rangle$ exhibits a pole in the variable
\begin{equation}
\alpha=\sum_{k=1}^m\alpha_k
\end{equation}
if the screening condition is satisfied
\begin{equation}
\alpha=Q-nb
\end{equation}
with a residue being expressed in terms of $2n$-dimensional Coulomb integral\footnote{The integrand in Eq \eqref{GL}  represents the expectation value of the fields $V_{\alpha_k}$ with $n$ screening fields $V_b(t_i)$ over the massless free-field vacuum state.}. Namely,
\begin{equation}\label{GL}
  \underset{\alpha=Q-nb}{\text{res}}\langle V_{\alpha_1}(z_1)\dots
  V_{\alpha_m}(z_m)\rangle=(-\pi\mu)^n\prod_{i<j}|z_i-z_j|^{-4\alpha_i\alpha_j}
  \int\prod_{j=1}^n\prod_{k=1}^m|t_j-z_k|^{-4b\alpha_k}\mathcal{D}_n^{-2b^2}(t)d^2\vec{t}_n,
\end{equation}
where $\mathcal{D}_n(t)$ is modulus square of Vandermonde determinant
\begin{equation}
\mathcal{D}_n(t)=\prod_{i<j}^n|t_i-t_j|^{2}.
\end{equation}
Throughout this paper we use the notation for measure
\begin{equation}
 d^2\vec{t}_n=\frac{1}{\pi^nn!}\prod_{k=1}^nd^2t_k.
\end{equation}
In the case of three points, integral in the r.~h.~s. of \eqref{GL} reads\footnote{Using projective invariance, one can set $z_1=0$, $z_2=1$ and $z_3=\infty$ in Eq \eqref{GL}.}
\begin{equation}\label{DF}
   I_n(\alpha_1,\alpha_2,\alpha_3)=
   \int\prod_{k=1}^n|t_k|^{-4b\alpha_1}|t_k-1|^{-4b\alpha_2}\mathcal{D}^{-2b^2}_n(t)\,d^2\vec{t}_n.
\end{equation}
Integral \eqref{DF} was calculated exactly in \cite{Dotsenko:1984ad}. Here we give the different derivation (simplest to our knowledge), based on the integral relation \cite{Baseilhac:1998eq}, which will be useful in the following 
\begin{multline}\label{Bazeyaka}
   \int\mathcal{D}_n(y)\prod_{i=1}^n\prod_{j=1}^{n+m+1}|y_i-t_j|^{2p_j}
   \,d^2\vec{y}_n=\\=
   \frac{\prod_j\gamma(1+p_j)}{\gamma(1+n+\sum_jp_j)}
   \prod_{i<j}|t_i-t_j|^{2+2p_i+2p_j}
   \int\mathcal{D}_m(u)\prod_{i=1}^m\prod_{j=1}^{n+m+1}|u_i-t_j|^{-2-2p_j}
   \,d^2\vec{u}_m.
\end{multline}
Namely, one should apply identity \eqref{Bazeyaka} for $m=0$ from the right to the left and substitute
\begin{equation}\label{Start_Bazeyaka}
  \mathcal{D}^{-2b^2}_n(t)=\mathcal{D}_n(t)
  \frac{\gamma(-nb^2)}{\gamma^n(-b^2)}
  \int\mathcal{D}_{n-1}(y)\prod_{i=1}^{n-1}\prod_{j=1}^{n}|y_i-t_j|^{-2-2b^2}d^2\vec{y}_{n-1}
\end{equation}
in r.~h.~s. of Eq \eqref{DF}. After that, integral over variables $t_j$ can be again calculated using Eq \eqref{Bazeyaka}. The remaining integral over $y_i$ will be of the same type as \eqref{DF}, but with lower number of integrations. As result, we derive the functional relation
\begin{equation}\label{Selberg_funk_Relat}
   I_n(\alpha_1,\alpha_2,\alpha_3)=\left(\frac{\gamma(-nb^2)}{\gamma(-b^2)}\right)
   \frac{1}
   {\gamma(2b\alpha_1)\gamma(2b\alpha_2)\gamma(2b\alpha_3+(n-1)b^2)}\;
   I_{n-1}(\alpha_1+b/2,\alpha_2+b/2,\alpha_3).
\end{equation}
Repeating this procedure $n$ times, we obtain that
\begin{equation}\label{Selberg}
   I_n(\alpha_1,\alpha_2,\alpha_3)=
   \prod_{k=1}^n\left(\frac{\gamma(-kb^2)}{\gamma(-b^2)}\right)
   \prod_{j=0}^{n-1}\frac{1}{\gamma(2b\alpha_1+jb^2)\gamma(2b\alpha_2+jb^2)
   \gamma(2b\alpha_3+jb^2)}.
\end{equation}
Functional relation \eqref{Selberg_funk_Relat} can be used to continue $I_n(\alpha_1,\alpha_2,\alpha_3)$ to the non-integer $n$ (the number of screenings $V_b$). Namely, one can substitute 
\begin{equation}\label{Src_cond}
n=(Q-\alpha_1-\alpha_2-\alpha_3)/b
\end{equation}
into Eq \eqref{Selberg_funk_Relat} and consider it, as a functional relation for the three-point function
$C(\alpha_1,\alpha_2,\alpha_3)$, which satisfies the condition that
\begin{equation}\label{C_res}
   \underset{\alpha=Q-nb}{\text{res}}C(\alpha_1,\alpha_2,\alpha_3)\overset{\text{def}}{=}
   (-\pi\mu)^n
   I_n(\alpha_1,\alpha_2,\alpha_3).
\end{equation}
Functional relation \eqref{Selberg_funk_Relat}, where $n$ is substituted by Eq \eqref{Src_cond}, together with residue condition \eqref{C_res} allow us to find this quantity.
An analytical expression for $C(\alpha_1,\alpha_2,\alpha_3)$ was proposed in \cite{Dorn:1992at,Dorn:1994xn,Zamolodchikov:1995aa} (see also \cite{Teschner:2003en})
\begin{equation}\label{3point}
    C(\alpha_1,\alpha_2,\alpha_3)=
    \Bigl[\pi\mu\gamma(b^2)b^{2-2b^2}\Bigr]^{\frac{(Q-\alpha)}{b}}
    \frac{\Upsilon'(0)\prod_{k=1}^3\Upsilon(2\alpha_k)}
    {\Upsilon(\alpha-Q)\prod_{k=1}^3\Upsilon(\alpha-2\alpha_k)},
\end{equation}
here $\Upsilon(x)$ is entire selfdual with respect to $b\rightarrow b^{-1}$ function satisfying functional relations 
\begin{equation}
 \begin{aligned}
   &\Upsilon(x+b)=\gamma(bx)b^{1-2bx}\Upsilon(x)\\
   &\Upsilon(x+b^{-1})=\gamma(b^{-1}x)b^{2b^{-1}x-1}\Upsilon(x)
 \end{aligned}
\end{equation}
and defined by the integral
\begin{equation}
    \log\Upsilon(x)=\int_{0}^{\infty}\frac{dt}{t}
    \Bigl[\left(\frac{Q}{2}-x\right)^2e^{-t}-\frac
    {\sinh^2\left(\frac{Q}{2}-x\right)\frac{t}{2}}
    {\sinh\frac{bt}{2}\sinh\frac{t}{2b}} 
    \Bigr].
\end{equation} 
This function is symmetric $\Upsilon(x)=\Upsilon(Q-x)$ and has zeros in points
\begin{equation}
    x=\begin{cases}
        -mb-nb^{-1},\\
        Q+mb+nb^{-1}, 
      \end{cases}\;\;
     m,n=0,1,2,\dots
\end{equation}
It is easy to see, that function $C(\alpha_1,\alpha_2,\alpha_3)$ satisfies relation \eqref{Selberg_funk_Relat}, where $n$ is defined by Eq \eqref{Src_cond}.

In the case of four points, the correlation function has much more complicated analytical structure. The situation simplifies, however, in the case, when one of the parameters $\alpha_k=-mb/2$ with \hbox{$m=0,1,2,\dots$} For these values of parameters, field $V_{-mb/2}(z)$ is degenerate at the level $m+1$ and four-point correlation function satisfies differential equation of the same  order in each variable $z$ and $\bar{z}$. It gives a reason to expect, that it can be represented by finite-dimensional integral. To find this integral representation we start from the case when the condition
\begin{equation}\label{alpha}
 -mb/2+\alpha_1+\alpha_2+\alpha_3+nb=Q
\end{equation}
is satisfied, i.~e. the correlation function possesses a pole in the variable $\alpha=\alpha_1+\alpha_2+\alpha_3$ with the residue
\begin{equation}
  \underset{\alpha=Q-nb+mb/2}{\text{res}}\langle   
   V_{-\frac{mb}{2}}(z)V_{\alpha_1}(0)V_{\alpha_2}(1)V_{\alpha_3}(\infty)\rangle
  =(-\pi\mu)^n|z|^{2mb\alpha_1}|z-1|^{2mb\alpha_2}J_m^{(n)}(\alpha_1,\alpha_2,\alpha_3|z)
\end{equation}
with
\begin{equation}\label{J_m^n}
   J_m^{(n)}(\alpha_1,\alpha_2,\alpha_3|z)=
  \int\prod_{k=1}^n|t_k|^{-4b\alpha_1}|t_k-1|^{-4b\alpha_2}
   |t_k-z|^{2mb^2}\mathcal{D}^{-2b^2}_n(t)d^2\vec{t}_n,
\end{equation}
here $n$ is the number of screening fields $V_b$.
One can reduce the number of integrations in Eq \eqref{J_m^n} by multiple application of relation \eqref{Bazeyaka}\footnote{We suppose for a moment, that $n\geq m$.}. Namely, it can be shown \cite{Fateev:2006}, that integral \eqref{J_m^n} satisfies the following remarkable property
\begin{equation}\label{4point-relation}
   J_m^{(n)}(\alpha_1,\alpha_2,\alpha_3|z)
    =\Omega_m^n(\alpha_1,\alpha_2,\alpha_3)
   J_n^{(m)}(\tilde{\alpha}_1,\tilde{\alpha}_2,\tilde{\alpha}_3|z),
\end{equation}
with $ \tilde{\alpha}_k=\alpha_k+(n-m)b/2$ and
\begin{equation}
  \Omega_m^n(\alpha_1,\alpha_2,\alpha_3)=
  \prod_{k=1}^{n-m}
  \left(\frac{\gamma(-(k+m)b^2)}{\gamma(-b^2)}\right)
  \prod_{j=0}^{n-m-1}\frac{1}{\gamma(2b\alpha_1+jb^2)\gamma(2b\alpha_2+jb^2)
   \gamma(2b\alpha_3+jb^2)}.
\end{equation}
Now, $n$  appears in the integral $J_n^{(m)}(\tilde{\alpha}_1,\tilde{\alpha}_2,\tilde{\alpha}_3|z)$ as  a parameter and we can easily perform continuation to the non-integer $n$ related with $\alpha$ by Eq \eqref{alpha}. Continuation of the factor $\Omega_m^n(\alpha_1,\alpha_2,\alpha_3)$ should be done in a such a way, that
\begin{equation*}
   \underset{\alpha=Q-nb+mb/2}{\text{res}}\Omega_m(\alpha_1,\alpha_2,\alpha_3)=
   (-\pi\mu)^n\Omega_m^n(\alpha_1,\alpha_2,\alpha_3).
\end{equation*}
As a result, we obtain the following expression for the four-point function with one degenerate field in LFT
\begin{equation}\label{Liouville_4point}
    \langle V_{-\frac{mb}{2}}(z)V_{\alpha_1}(0)V_{\alpha_2}(1)V_{\alpha_3}(\infty)\rangle=
    \Omega_m(\alpha_1,\alpha_2,\alpha_3)\;\vert z\vert^{2mb\alpha_1}
  \vert z-1\vert^{2mb\alpha_2}\,\mathbf{J}_m(A,B,C|z),
\end{equation}
\begin{equation}\label{4point}
  \mathbf{J}_m(A,B,C|z)=
  \int\prod_{k=1}^m
  \vert t_k\vert^{2A}\vert t_k-1\vert^{2B}
  \vert t_k-z\vert^{2C}\mathcal{D}^{-2b^2}_m(t)\,d^2\vec{t}_m
\end{equation}
with
\begin{equation*}
       A=b\left(\alpha-2\alpha_1-Q+mb/2\right),\;\;
       B=b\left(\alpha-2\alpha_2-Q+mb/2\right),\;\;
       C=b\left(Q+mb/2-\alpha\right)
\end{equation*}
and normalization constant $\Omega_m(\alpha_1,\alpha_2,\alpha_3)$ is
\begin{equation}\label{O_m}
   \Omega_m(\alpha_1,\alpha_2,\alpha_3)=
   (-\pi\mu)^m
   \Bigl[\pi\mu\gamma(b^2)b^{2-2b^2}\Bigr]^{\frac{(Q-\alpha-mb/2)}{b}}\;
   \frac{\Upsilon'(-mb)\prod_{k=1}^3\Upsilon(2\alpha_k)}
    {\Upsilon(\alpha-Q-\frac{mb}{2})\prod_{k=1}^3\Upsilon(\alpha-2\alpha_k+\frac{mb}{2})},
\end{equation}
here $\alpha=\alpha_1+\alpha_2+\alpha_3$.

It is possible to generalize results \eqref{3point} and \eqref{4point} for multipoint correlation functions. For example, for five-point correlation function with two degenerate fields 
\begin{equation*}
 \langle V_{-\frac{m_1b}{2}}(z_1)
  V_{-\frac{m_2b}{2}}(z_2)V_{\alpha_1}(0)V_{\alpha_2}(1)V_{\alpha_3}(\infty)\rangle
\end{equation*}
one can obtain the following expression
\begin{multline}\label{5point}
  \langle V_{-\frac{m_1b}{2}}(z_1)
  V_{-\frac{m_2b}{2}}(z_2)V_{\alpha_1}(0)V_{\alpha_2}(1)V_{\alpha_3}(\infty)\rangle
   =\Omega_{m_1m_2}(\alpha_1,\alpha_2,\alpha_3)
  |z_1|^{2m_1b\alpha_1}\,|z_1-1|^{2m_1b\alpha_2}\times\\\times
  |z_2|^{2m_2b(Q-\alpha_1)}\,|z_2-1|^{2m_2b(Q-\alpha_2)}\;|z_1-z_2|^{-m_1m_2b^2}
  \mathbf{J}_{m_1m_2}(A,B,C|z_1,z_2),
\end{multline}
where
\begin{multline}\label{Jmm}
  \mathbf{J}_{m_1m_2}(A,B,C|z_1,z_2)=
  \int\mathcal{D}^{-2b^2}_{m_1}(t)\;\mathcal{D}^{-2b^2}_{m_2}(y)
  \prod_{j=1}^{m_1}\prod_{k=1}^{m_2}|t_j-y_k|^{2b^2}
  \prod_{j=1}^{m_1}|t_j|^{2A}\,|t_j-1|^{2B}\,|t_j-z_1|^{2C}\,d^2\vec{t}_{m_1}\times\\\times
  \prod_{k=1}^{m_2}|y_k|^{-2(1+b^2+A)}\,|y_k-1|^{-2(1+b^2+B)}\,|y_k-z_2|^{-2(1-(m_1+m_2-1)b^2+C)}
  \,d^2\vec{y}_{m_2},
\end{multline}
with
\begin{equation*}
  A=b\bigl(\alpha-2\alpha_1-Q+\frac{(m_1-m_2)b}{2}\bigr);\;
  B=b\bigl(\alpha-2\alpha_2-Q+\frac{(m_1-m_2)b}{2}\bigr);\;
  C=b\bigl(Q+\frac{(m_1+m_2)b}{2}-\alpha\bigr)
\end{equation*}
and normalization factor $\Omega_{m_1m_2}(\alpha_1,\alpha_2,\alpha_3)$ is equal
\begin{multline}
 \Omega_{m_1m_2}(\alpha_1,\alpha_2,\alpha_3)=
   (-\pi\mu)^{m_1+m_2}\;
   \Bigl[\pi\mu\gamma(b^2)b^{2-2b^2}\Bigr]^{\frac{(Q-\alpha-(m_1+m_2)b/2)}{b}}\;
   \frac{\Upsilon'(-m_1b)\Upsilon'(-m_2b)}{\Upsilon'(0)}\times\\\times
   \frac{\Upsilon\Bigl(\alpha-Q+\frac{(m_1-m_2)b}{2}\Bigr)}
   {\Upsilon\Bigl(\alpha-Q+\frac{(m_1+m_2)b}{2}\Bigr)\Upsilon\Bigl(\alpha-Q-\frac{(m_1+m_2)b}{2}\Bigr)}\;
    \prod_{k=1}^3\frac{\Upsilon(2\alpha_k)}
    {\Upsilon\Bigl(\alpha-2\alpha_k+\frac{(m_1-m_2)b}{2}\Bigr)}.
\end{multline}
We note that the integral for five-point correlation function  \eqref{Jmm} has a hidden $sl(3)$ structure. Namely, Coulomb interaction  between point $t_i$ and $y_j$ in the first three terms in \eqref{Jmm} can be described by Cartan matrix of the Lie algebra $sl(3)$. The integrals with that type of Coulomb interaction appear in conformal Toda field theory \cite{Fateev:2005gs}. Similar $sl(m+1)$ structure appears in the Coulomb integrals for $m+3$-point correlation function with three arbitrary fields.

In principle, it is possible to find analytical expression for any $k+3$-point correlation function with $k$ degenerate and three arbitrary fields. For these purpose it is sufficient to write down the integral representation for the correlation function with $m$ fields $V_{-\frac{b}{2}}$ 
\begin{equation}\label{npoint-def}
  \langle V_{-\frac{b}{2}}(z_1)\dots
  V_{-\frac{b}{2}}(z_m)V_{\alpha_1}(0)V_{\alpha_2}(1)V_{\alpha_3}(\infty)\rangle,
\end{equation}
because correlation function with arbitrary degenerate fields can be obtained from this function using  the fusion property
\begin{equation}
  V_{-\frac{m_1b}{2}}(z)V_{-\frac{m_2b}{2}}(0)=|z|^{-m_1m_2b^2}V_{-\frac{(m_1+m_2)b}{2}}(0)+\dots
\end{equation}
The explicit expression for the correlation function \eqref{npoint-def} can be written in terms of auxiliary function (kernel) $K_m^{\Delta}(t_1,..,t_m|y_1,..,y_m)$, which is symmetric function of variables $t_k$ and $y_k$ and does not change under the permutation $t_k\leftrightarrow y_k$. The last two properties are not evident from the explicit form of the function $K_m^{\Delta}(t_1,..,t_m|y_1,..,y_m)$, but they can be proved. This function is given by the $m(m-1)$-dimensional Coulomb integral and can be derived from the recurrent formula\footnote{To make sense of this formula we set $K_0^{\Delta}=1$.}
\begin{multline}\label{Kernel}
   K_m^{\Delta}(t_1,..,t_m|y_1,..,y_m)=
   \frac{\gamma(-mb^2)}{\gamma^m(-b^2)}\,\mathcal{D}_m^{1+2b^2}(t)\;
   \prod_{k=1}^m|t_k-y_1|^{2\Delta}
   \times\\\times
   \int\mathcal{D}_{m-1}(\tau)\prod_{j=1}^{m-1}|\tau_j-y_1|^{-2\Delta+2b^2}
   \prod_{k=1}^{m}|\tau_j-t_k|^{-2-2b^2}
   K_{m-1}^{\Delta+b^2}(\tau_1,..,\tau_{m-1}|y_2,..,y_m)\,d^2\vec{\tau}_{m-1}.
\end{multline}
We give the derivation of this function for the cases $m=1$ and $m=2$  and describe the general procedure for  $m>2$ in the Appendix \ref{App-Kernel}. 

Being rather complicated this function has simple semiclassical asymptotic at $b\rightarrow0$ while $\Delta$ keeping fixed
\begin{equation}
  K_m^{\Delta}(t_1,..,t_m|y_1,..,y_m)\underset{b\rightarrow0}{\longrightarrow}\frac{1}{m!}
  \Bigl[|t_1-y_1|^{2\Delta}|t_2-y_2|^{2\Delta}\dots|t_m-y_m|^{2\Delta}+
  \text{symmetrization}\Bigr].
\end{equation}
In order to clarify Eq \eqref{Kernel} we give first three examples. Function $K_1^{\Delta}(t|y)$ is the simplest one
\begin{equation}\label{K1}
  K_1^{\Delta}(t|y)=|t-y|^{2\Delta}.
\end{equation}
Function $K_2^{\Delta}(t_1,t_2|y_1,y_2)$ is given by two-dimensional integral
\begin{multline}\label{K2}
  K_2^{\Delta}(t_1,t_2|y_1,y_2)=\frac{\gamma(-2b^2)}{\gamma^2(-b^2)}\,
  |t_1-t_2|^{2+4b^2}\,\,|t_1-y_1|^{2\Delta}|t_2-y_1|^{2\Delta}\times\\\times
  \int|\xi-y_1|^{-2\Delta+2b^2}|\xi-y_2|^{2\Delta+2b^2}|\xi-t_1|^{-2-2b^2}|\xi-t_2|^{-2-2b^2}
   \,d^2\vec{\xi}_1.
\end{multline}
Symmetry properties of the kernel \eqref{K2} can be verified using integral relation \eqref{App-int-4point}. Function $K_3^{\Delta}(t_1,t_2,t_3|y_1,y_2,y_3)$ is given by six-dimensional Coulomb integral
\begin{multline}
  K_3^{\Delta}(t_1,t_2,t_3|y_1,y_2,y_3)=\frac{\gamma(-2b^2)\gamma(-3b^2)}{\gamma^5(-b^2)}
 \mathcal{D}_3^{1+2b^2}(t)\prod_{k=1}^3|t_k-y_1|^{2\Delta}\times\\\times
 \int \mathcal{D}_2^{2+2b^2}(\nu)\prod_{k=1}^2|\nu_k-y_1|^{-2\Delta+2b^2}|\nu_k-y_2|^{2\Delta+2b^2}
  \prod_{j=1}^3|\nu_k-t_j|^{-2-2b^2}\,d^2\vec{\nu}_2\\
 |\xi-y_2|^{-2\Delta}|\xi-y_3|^{2\Delta+4b^2}
  |\xi-\nu_1|^{-2-2b^2}|\xi-\nu_2|^{-2-2b^2}\,d^2\vec{\xi}_1
\end{multline}
and so on. 

Correlation function \eqref{npoint-def} can be expressed in terms of the kernel \eqref{Kernel} as:
\begin{multline}\label{n-point}
\langle V_{-\frac{b}{2}}(z_1)\dots
  V_{-\frac{b}{2}}(z_m)V_{\alpha_1}(0)V_{\alpha_2}(1)V_{\alpha_3}(\infty)\rangle=
  \Omega_m(\alpha_1,\alpha_2,\alpha_3)\prod_{k=1}^m\vert z_k\vert^{2b\alpha_1}
  \vert z_k-1\vert^{2b\alpha_2}\times\\\times
  \prod_{i<j}|z_i-z_j|^{-b^2}
  \int\prod_{k=1}^m\vert t_k\vert^{2A}\vert t_k-1\vert^{2B}K_m^C(t_1,..,t_m|z_1,..,z_m)
  \mathcal{D}_m^{-2b^2}(t)\,d^2\vec{t}_m
\end{multline}
where
\begin{equation}
       A=b\left(\alpha-2\alpha_1-Q+mb/2\right),\;\;
       B=b\left(\alpha-2\alpha_2-Q+mb/2\right),\;\;
       C=b\left(Q+(2-m)b/2-\alpha\right),
\end{equation}
and normalization factor $\Omega_m(\alpha_1,\alpha_2,\alpha_3)$ is given by Eq \eqref{O_m}. 

Kernel \eqref{Kernel} despite its complicated form has many remarkable properties, which allow to obtain correlation function for higher degenerate fields $V_{-mb/2}$. For example it has very simple asymptotic at the limit $y_k\rightarrow z$
\begin{equation}\label{y_k=y}
  K_m^{\Delta}(t_1,..,t_m|z,..,z)=\prod_{k=1}^m|t_k-z|^{2\Delta+2(m-1)b^2}.
\end{equation}
Using \eqref{y_k=y} we can obtain from Eq \eqref{n-point} expression for the four-point function \eqref{Liouville_4point}. Another important property of the kernel \eqref{Kernel} is given by Eq \eqref{Kernel-razbienie}. If we consider $2m$-dimensional integral \eqref{n-point} with $m=m_1+m_2$ and "factorise" it using \eqref{Kernel-razbienie} to $2m_1$ and $2m_2$ dimensional integrals; then applying Eq \eqref{y_k=y} we derive the expression for the five-point correlation function \eqref{5point}. 
\section{Application: minimal Liouville gravity}\label{MLG}
In this section we consider minimal Liouville gravity, where the matter part of theory is described by generalized minimal model (GMM) of CFT \cite{Zamolodchikov:2005fy} with central charge
\begin{equation}
  c_M=1-6\bigl(b-b^{-1}\bigr)^2,
\end{equation}
coupled to the LFT in such a way, that $c_L+c_M=26$. GMM theory includes a continuous number of primary fields $\Phi_{\alpha}$ with conformal dimensions $\Delta_M(\alpha)=\alpha(\alpha+b-b^{-1})$. Multipoint correlation functions of these fields posses integral representation if the screening condition for matter fields is satisfied, i.~e.
\begin{equation}\label{Matter-Scr}
  \sum_{k=1}^m\alpha_k-nb=b^{-1}-b,
\end{equation}
then\footnote{We note that central charge $c_M$, conformal dimension $\Delta_M(\alpha)$, screening condition \eqref{Matter-Scr} and correlation function \eqref{Matter_npoint} (up to normalization) can be derived from the corresponding values in LFT by the substitution $b\rightarrow-ib$ and $\alpha_k\rightarrow i\alpha_k$.}
\begin{equation}\label{Matter_npoint}
  \langle\Phi_{\alpha_1}(z_1)\dots\Phi_{\alpha_m}(z_m)\rangle=\frac{1}{N(b^{-1}-b)}
  \prod_{k=1}^mN(\alpha_k)\prod_{i<j}|z_i-z_j|^{4\alpha_i\alpha_j}
  \int\prod_{j=1}^n\prod_{k=1}^m|t_j-z_k|^{-4b\alpha_k}\mathcal{D}_n^{2b^2}(t)d^2\vec{t}_n
\end{equation}
where the normalization factor
\begin{equation}
  N(\alpha)=\bigl(-\gamma(-b^2)\bigr)^{-\frac{\alpha}{b}}
  \left[\frac{\gamma(2b^{-1}\alpha-b^{-2}+2)}{\gamma(2b\alpha+b^2)}
  \frac{\gamma(b^{2})}{\gamma(2-b^{-2})}\right]^{\frac{1}{2}}
\end{equation}
is chosen in such a way that
\begin{equation}
  \langle\Phi_{\alpha}(z)\Phi_{\alpha}(z')\rangle=|z-z'|^{-4\Delta_M(\alpha)}.
\end{equation}

One of the main problems in minimal Liouville gravity is to calculate correlation functions of operators 
\begin{equation}
  U_{\alpha}=\Phi_{\alpha-b}V_{\alpha},
\end{equation}
which have conformal dimension $\Delta_L(\alpha)+\Delta_M(\alpha-b)=1$ and hence $(1,1)$ form $U_{\alpha}(z)d^2z$ can be integrated in an invariant way. Integrated $n$-point correlation functions are invariant objects, which depend only on parameters $\alpha_k$. Due to the group of diffeomorphism $SL(2,C)$, which is the symmetry of the theory, the number of integrations in $m+3$--point correlation function can be reduced to $m$. Namely, we can fix the coordinates of any three fields at the points $0$, $1$ and $\infty$. In this way correlation function (or more correctly correlation number) of integrated operators $U_{\alpha}$ can be defined as
\begin{multline}\label{Gravity_definition}
  \langle\langle U_{\alpha_1}\dots U_{\alpha_{m+3}}\rangle\rangle
  \overset{\text{def}}{=}\frac{1}{Z_L}\int
  \langle V_{\alpha_1}(z_1)\dots V_{\alpha_{m}}(z_m)
   V_{\alpha_{m+1}}(0)V_{\alpha_{m+2}}(1)V_{\alpha_{m+3}}(\infty)\rangle\times\\\times
  \langle \Phi_{\alpha_1-b}(z_1)\dots \Phi_{\alpha_{m}-b}(z_m)
   \Phi_{\alpha_{m+1}-b}(0)\Phi_{\alpha_{m+2}-b}(1)\Phi_{\alpha_{m+3}-b}(\infty)\rangle\;
   d^2z_1\dots d^2z_m,
\end{multline}
where $Z_L$ is Liouville partition function \cite{Zamolodchikov:2005fy}
\begin{equation}
  Z_L=\bigl(\pi\mu\gamma(b^2)\bigr)^{Q/b}\frac{(1-b^2)}{\pi^3\gamma(b^2)\gamma(b^{-2})Q}.
\end{equation}

Three-point correlation function of the fields $U_{\alpha_k}$ does not contain integration and has simple factorized form \cite{Zamolodchikov:2005fy}
\begin{equation}\label{Gr-3point}
   \langle\langle U_{\alpha_1}U_{\alpha_2}U_{\alpha_3}\rangle\rangle=b^{-2}(b^{-4}-1) 
   \prod_{k=1}^3\mathcal{N}(\alpha_k),
\end{equation}
where we have introduced so called leg-factor \cite{Belavin:2005yj,Belavin:2006ex}
\begin{equation}
 \mathcal{N}(\alpha)=\pi\bigl(\pi\mu\gamma(b^2)\bigr)^{-\alpha/b}
 \left[\frac{\gamma(2b\alpha-b^2)\gamma(2b^{-1}\alpha-b^{-2})}
  {\gamma(1-b^2)\gamma(2-b^{-2})}\right]^{\frac{1}{2}}.
\end{equation}

Four-point function \eqref{Gravity_definition} contains one integration over $z$. We consider particular four-point function
\begin{equation}\label{Gravity_4point}
 \langle\langle U_{-\frac{mb}{2}}U_{\alpha_1}U_{\alpha_2}U_{\alpha_3}\rangle\rangle,
\end{equation}
which contains one degenerate Liouville field $V_{-\frac{mb}{2}}$ and hence Liouville part of correlation function is rather simple and is given by the integral \eqref{4point}. However matter part of this correlation function remains complicated. To simplify it we suppose, that in matter sector the screening condition is satisfied
\begin{equation}\label{4point-scr}
  -mb/2+\alpha_1+\alpha_2+\alpha_3-4b-nb=b^{-1}-b.
\end{equation}
In this case due to Eq \eqref{Matter_npoint} matter part of the correlation function is given in terms of $2n$ dimensional Coulomb integral. Under these two assumptions correlation function \eqref{Gravity_4point} can be expressed in terms of $2(n+m+1)$ dimensional integral (modulo numerical factors, which come from Eqs \eqref{Liouville_4point} and \eqref{Matter_npoint})
\begin{multline}\label{Gravity-4point}
   \mathbf{G}_{m}^n(\alpha_1,\alpha_2,\alpha_3)=
  \int\mathcal{D}^{-2b^2}_m(t)\;\mathcal{D}^{2b^2}_n(\xi)
  \prod_{j=1}^{m}|t_j|^{2(m+n+2)b^2-4b\alpha_1}\,|t_j-1|^{2(m+n+2)b^2-4b\alpha_2}\,
  |t_j-z|^{-2(n+2)b^2}\times\\\times
  |z|^{2(m+2)b^2-4b\alpha_1}\,|z-1|^{2(m+2)b^2-4b\alpha_2}
  |\xi_j|^{-4b(\alpha_1-b)}|\xi_j-1|^{-4b(\alpha_2-b)}|\xi_j-z|^{2(m+2)b^2}
  d^2\vec{t}_m\,d^2\vec{\xi}_n\,d^2z.
\end{multline}
Integral \eqref{Gravity-4point} has very simple structure of poles in each of the variables $\alpha_i$ and can be calculated exactly (see Ref \cite{Fateev:2006}) with the result
\begin{multline}
  \mathbf{G}_{m}^n(\alpha_1,\alpha_2,\alpha_3)=\pi(m+1)\,(n+1)\;
  \prod_{j=1}^{m+1}\left(\frac{\gamma(-jb^2)}{\gamma(-b^2)}\right)
  \prod_{j=1}^{n+1}\left(\frac{\gamma(jb^2)}{\gamma(b^2)}\right)\times\\\times
  \prod_{j=2}^{m+n+2}
  \left[\gamma(2b\alpha_1-jb^2)\gamma(2b\alpha_2-jb^2)\gamma(2b\alpha_3-jb^2)\right]^{-1}.
\end{multline}
Taking together all normalization factors coming from Eqs \eqref{Liouville_4point} and \eqref{Matter_npoint} we arrive at a final simple result
\begin{equation}\label{Gr-4point}
  \langle\langle U_{-\frac{mb}{2}}U_{\alpha_1}U_{\alpha_2}U_{\alpha_3}\rangle\rangle
  =b^{-2}(b^{-4}-1)
  \prod_{k=1}^4\mathcal{N}(\alpha_k)\,(m+1)(n+1),
\end{equation}
where $\alpha_4=-mb/2$. Formula \eqref{Gr-4point} is valid in the domain of convergency of the integral \eqref{Gravity-4point}
\begin{equation}
  Q+b<\sum_{k=1}^4\alpha_k<2Q-mb;\;\;(m+1)b/2<\alpha_j<Q/2\;\;\;j=1,2,3.
\end{equation}
In this domain one can express the number of screenings $n$ from Eq \eqref{4point-scr}.

Five-point function \eqref{Gravity_definition} will have two integration over variables $z_1$ and $z_2$. We consider the particular correlation function, which contains two degenerate Liouville fields
\begin{equation}\label{Gravity_5point}
 \langle\langle U_{-\frac{m_1b}{2}}U_{-\frac{m_2b}{2}}U_{\alpha_1}U_{\alpha_2}U_{\alpha_3}\rangle\rangle
\end{equation}
and assume again, that in the matter sector we have screening condition
\begin{equation}\label{5point-scr}
  -m_1b/2-m_2b/2+\alpha_1+\alpha_2+\alpha_3-5b-nb=b^{-1}-b.
\end{equation}
In this case correlation function \eqref{Gravity_5point} can be expressed in terms of $2(m_1+m_2+n+2)$ dimensional integral (modulo numerical factors, which come from Eqs \eqref{5point} and \eqref{Matter_npoint})
\begin{multline}\label{Gravity-5point}
   \mathbf{G}_{m_1m_2}^n(\alpha_1,\alpha_2,\alpha_3)=\int
   \mathbf{J}_{m_1m_2}(A,B,C|z_1,z_2)\;
  |z_1|^{2(m_1+2)b^2-4b\alpha_1}\,|z_1-1|^{2(m_1+2)b^2-4b\alpha_2}\times\\\times
  |z_2|^{-2+2(m_2+1)(1+2b^2-2b\alpha_1)}\,|z_2-1|^{-2+2(m_2+1)(1+2b^2-2b\alpha_2)}
  |z_1-z_2|^{2(m_1+m_2+2)b^2}\\\prod_{j=1}^n
  |\xi_j|^{-4b(\alpha_1-b)}|\xi_j-1|^{-4b(\alpha_2-b)}|\xi_j-z_1|^{2(m_1+2)b^2}|\xi_j-z_2|^{2(m_2+2)b^2}
  \mathcal{D}^{2b^2}_n(\xi)d^2\vec{\xi}_{n}\,d^2z_1\,d^2z_2,
\end{multline}
where function $\mathbf{J}_{m_1m_2}(A,B,C|z_1,z_2)$ is given by Eq \eqref{Jmm} with
\begin{equation*}
A=(m_1+n+3)b^2-2b\alpha_1,\;B=(m_1+n+3)b^2-2b\alpha_2,\;C=-(n+3)b^2.
\end{equation*}
This integral as a function of the parameters $\alpha_i$ also has simple structure of poles and can be calculated exactly with the result
\begin{multline}
  \mathbf{G}_{m_1m_2}^n(\alpha_1,\alpha_2,\alpha_3)=
  \pi^2(m_1+1)\,(m_2+1)\,(n+1)(n+2)
  \prod_{j=1}^{m_1+1}\left(\frac{\gamma(-jb^2)}{\gamma(-b^2)}\right)
  \prod_{j=1}^{m_2+1}\left(\frac{\gamma(-jb^2)}{\gamma(-b^2)}\right)
  \prod_{j=1}^{n+2}\left(\frac{\gamma(jb^2)}{\gamma(b^2)}\right)\\\times
  \prod_{j=1}^{m_2}\gamma\bigl((m_1+n+2+j)b^2\bigr)\;
  \prod_{j=2}^{m_1+n+3}
  \left[\gamma(2b\alpha_1-jb^2)\gamma(2b\alpha_2-jb^2)\gamma(2b\alpha_3-jb^2)\right]^{-1}.
\end{multline}
Collecting all needed factors one obtains a simple result for five-point correlation function
\begin{equation}\label{Gr-5point}
  \langle\langle U_{-\frac{m_1b}{2}}U_{-\frac{m_2b}{2}}U_{\alpha_1}U_{\alpha_2}U_{\alpha_3}
  \rangle\rangle=b^{-2}(b^{-4}-1)\prod_{k=1}^5\mathcal{N}(\alpha_k)\,(m_1+1)(m_2+1)(n+1)(n+2),
\end{equation}
where $\alpha_4=-m_1b/2$ and $\alpha_5=-m_2b/2$. Result \eqref{Gr-5point} is valid in the domain of convergency of the integral \eqref{Gravity-5point}
\begin{equation}
  Q+2b<\sum_{k=1}^5\alpha_k<2Q-(m_1+m_2)b;\;\;(m_1+m_2+1)b/2<\alpha_j<Q/2\;\;\;j=1,2,3.
\end{equation}

Eqs \eqref{Gr-3point}, \eqref{Gr-4point} and \eqref{Gr-5point} can be easily generalized. Let us consider $k+3$--point correlation function with $k$ degenerate Liouville fields
\begin{equation}\label{Gravity_npoint}
 \langle\langle U_{-\frac{m_1b}{2}}\dots U_{-\frac{m_kb}{2}}
  U_{\alpha_1}U_{\alpha_2}U_{\alpha_3}\rangle\rangle
\end{equation}
and assume, that in the matter sector we have screening condition (compare with Eqs \eqref{4point-scr} and \eqref{5point-scr})
\begin{equation}
  -m_1b/2-\dots-m_kb/2+\alpha_1+\alpha_2+\alpha_3-(k+3)b-nb=b^{-1}-b.
\end{equation}
It looks reasonable to expect, that there exist a domain in the space of parameters $\alpha_k$, where Eqs \eqref{Gr-3point}, \eqref{Gr-4point} and \eqref{Gr-5point} admit straitforward generalization
\begin{equation}\label{Gr-npoint}
  \langle\langle U_{-\frac{m_1b}{2}}\dots U_{-\frac{m_kb}{2}}
  U_{\alpha_1}U_{\alpha_2}U_{\alpha_3}\rangle\rangle=
  b^{-2}(b^{-4}-1)
  \prod_{j=1}^{k+3}\mathcal{N}(\alpha_j)\,\prod_{j=1}^k(m_j+1)(n+j),
\end{equation}
where we have set $\alpha_{j+3}=-m_jb/2$.

All results of this paper can be rewritten in the "dual" case, where we consider $k+3$-point correlation function with $k$ degenerate matter fields $\Phi_{m_jb/2}$, i.~e. correlator
\begin{equation}\label{Gravity_npoint-dual}
 \langle\langle U_{\frac{(m_1+2)b}{2}}\dots U_{\frac{(m_k+2)b}{2}}
  U_{\alpha_1}U_{\alpha_2}U_{\alpha_3}\rangle\rangle
\end{equation}
and assume that in the Liouville sector the screening condition is satisfied
\begin{equation}
  m_1b/2+\dots+m_kb/2+\alpha_1+\alpha_2+\alpha_3+kb+nb=b+b^{-1},
\end{equation}
i.~e. Liouville part of correlation function  \eqref{Gravity_npoint-dual} is given by $2n$-dimensional integral due to Eq \eqref{GL}. In this case the matter part of correlation function \eqref{Gravity_npoint-dual} 
\begin{equation}\label{Matter-npoint}
 \langle\Phi_{\frac{m_1b}{2}}\dots\Phi_{\frac{m_kb}{2}}
  \Phi_{\alpha_1-b}\Phi_{\alpha_2-b}\Phi_{\alpha_3-b}\rangle
\end{equation}
can be represented by the integrals, similar to those, which were derived in the previous section for LFT. Namely, integrals for $k+3$-point correlation function in GMM with $k$ degenerate field $\Phi_{m_jb/2}$ and three arbitrary fields $\Phi_{\alpha_j}$ can be obtained by substitution $\alpha_j\rightarrow i\alpha_j$ and $b\rightarrow-ib$ into integrals for $k+3$-point function in the LFT with $k$ degenerate and three arbitrary fields $V_{\alpha_j}$ (note that in correlation function \eqref{Matter-npoint} parameters $\alpha_j$ are shifted $\alpha_j\rightarrow\alpha_j-b$).  However overall numerical factor will be given by another expression. This factor for the case of four-point correlation function can be found in Ref \cite{Fateev:2006}, where four-point correlation function \eqref{Gravity_npoint-dual} were considered.
\subsection*{Acknowledgment}
Work of V.~F. was supported by the European Committee under contract EUCLID HRPN-CT-2002-00325.
A.~L. acknowledges the support by Russian Foundation for Basic Research under the grant RBRF 07-02-00799-a, by Russian Ministry of Science and Technology under the Scientific Schools grant 2044.2003.2 and by RAS program "Elementary particles and the fundamental nuclear physics". An important part has been made during his visit at the Laboratoire de Physique Th\'eorique et Astroparticules Universit\'e Montpellier~II within ENS-LANDAU program.
\Appendix
\section{Calculation of the kernel}\label{App-Kernel}
In this section we give some hints how to derive Eq \eqref{n-point}. The idea is completely analogous to that, which was used for derivation of three-point correlation function in section \ref{Liouville}. We start with correlation function
\begin{equation}\label{App-mpoint-def}
  \langle V_{-\frac{b}{2}}(z_1)\dots
  V_{-\frac{b}{2}}(z_m)V_{\alpha_1}(0)V_{\alpha_2}(1)V_{\alpha_3}(\infty)\rangle,
\end{equation}
and assume, that the screening condition is satisfied, i.~e.
\begin{equation}\label{App-scr}
  -\frac{mb}{2}+\alpha_1+\alpha_2+\alpha_3+nb=Q.
\end{equation}
As a consequence of Eq \eqref{GL} this correlation function is expressed in terms of $2n$-dimensional Coulomb integral 
\begin{equation}\label{App-mpoint}
  \mathfrak{J}_m^{(n)}(A,B|z_1,..,z_m)=
  \int\prod_{k=1}^n |t_k|^{2A}|t_k-1|^{2B}\prod_{j=1}^m|t_k-z_j|^{-2g}\mathcal{D}_n^{2g}(t)\,d^2\vec{t}_n,
\end{equation}
where we have set for shortness
\begin{equation}
  A=-2b\alpha_1,\;\;B=-2b\alpha_2,\;\;g=-b^2.
\end{equation}
The main idea is to apply identity \eqref{Bazeyaka} several times and reduce integral \eqref{App-mpoint} to another integral with number of integrations, which does not depend on the number of screenings $n$ (dependence on $n$ appears in this integral as a parameter). After that one can continue this integral to the non-integer value of $n$ expressing $n$  from Eq \eqref{App-scr}.

We show how does it work for the cases of four and five point correlation functions. 
Below in this appendix we use identity \eqref{Bazeyaka} many times and for shortness, we omit irrelevant factors from $\gamma$ functions\footnote{It is easily to restore these factors in the final answer from general principles.}. 
\paragraph*{Four-point function.}
In the case of four-point function we have integral
\begin{equation}\label{App-4point}
  \mathfrak{J}_1^{(n)}(A,B|z)=
  \int\prod_{k=1}^n |t_k|^{2A}|t_k-1|^{2B}|t_k-z|^{-2g}\mathcal{D}_n^{2g}(t)\,d^2\vec{t}_n.
\end{equation}
To start, we use identity \eqref{Bazeyaka} from the right to the left and substitute 
\begin{equation}\label{Bazeyaka-start}
  \mathcal{D}^{2g}_n(t)\sim\mathcal{D}_n(t)
  \int\mathcal{D}_{n-1}(y)\prod_{i=1}^{n-1}\prod_{j=1}^{n}|y_i-t_j|^{-2+2g}d^2\vec{y}_{n-1}
\end{equation}
into Eq \eqref{App-4point}. In Eq \eqref{Bazeyaka-start} and below sign $\sim$ means, that we have omitted irrelevant numerical factors (compare with Eq \eqref{Start_Bazeyaka}, where this factor is given explicitly). After that using again identity \eqref{Bazeyaka} 
we can perform integration over the variables $t_k$ and convert this integral to the $2$ dimensional integral over auxiliary variable $\xi$ 
\begin{multline}
  \int\prod_{k=1}^n |t_k|^{2A}|t_k-1|^{2B}|t_k-z|^{-2g}\prod_{j=1}^{n-1}|t_k-y_j|^{-2+2g}
  \mathcal{D}_n(t)d^2\vec{t}_n\sim|z|^{2(1+A-g)}|z-1|^{2(1+B-g)}
  \times\\\times
  \prod_{k=1}^{n-1}|y_k|^{2(A+g)}|y_k-1|^{2(B+g)}\mathcal{D}_{n-1}^{2g-1}(y)
  \int|\xi|^{-2-2A}|\xi-1|^{-2-2B}|\xi-z|^{-2+2g}\prod_{j=1}^{n-1}|\xi-y_j|^{-2g}\,d^2\xi.
\end{multline}
As a result we obtain the following relation between integrals $\mathfrak{J}_1^{(n)}$ and $\mathfrak{J}_1^{(n-1)}$
\begin{multline}\label{App-4point-recurent-1}
  \mathfrak{J}_1^{(n)}(A,B|z)\sim|z|^{2(1+A-g)}|z-1|^{2(1+B-g)}\times\\\times
  \int|\xi|^{-2(1+A)}|\xi-1|^{-2(1+B)}|\xi-z|^{-2+2g}\;\mathfrak{J}_1^{(n-1)}(A+g,B+g|\xi)\,d^2\xi.
\end{multline}
Applying relation \eqref{App-4point-recurent-1} again we obtain
\begin{multline}
  \mathfrak{J}_1^{(n)}(A,B|z)\sim|z|^{2(1+A-g)}|z-1|^{2(1+B-g)}\times\\\times
  \int|\xi-z|^{-2+2g}
  |\nu|^{-2(1+A+g)}|\nu-1|^{-2(1+B+g)}|\nu-\xi|^{-2+2g}\mathfrak{J}_1^{(n-2)}(A+2g,B+2g|\nu)
  \,d^2\xi\,d^2\nu.
\end{multline}
Integral over $d^2\xi$ can be easily performed and up to trivial factor is equal to $|\nu-z|^{-2+4g}$.
As a result we obtain relation between integrals $\mathfrak{J}_1^{(n)}$ and $\mathfrak{J}_1^{(n-2)}$
\begin{multline}\label{App-4point-recurent-2}
  \mathfrak{J}_1^{(n)}(A,B|z)\sim|z|^{2(1+A-g)}|z-1|^{2(1+B-g)}\times\\\times
  \int|\nu|^{-2(1+A+g)}|\nu-1|^{-2(1+B+g)}|\nu-z|^{-2+4g}\;\mathfrak{J}_1^{(n-2)}(A+2g,B+2g|\nu)\,d^2\nu.
\end{multline}
Equations \eqref{App-4point-recurent-1} and \eqref{App-4point-recurent-2} can be easily written after arbitrary number $k$ of applications of our procedure
\begin{multline}\label{App-4point-recurent-k}
  \mathfrak{J}_1^{(n)}(A,B|z)\sim|z|^{2(1+A-g)}|z-1|^{2(1+B-g)}\times\\\times
  \int|\nu|^{-2(1+A+(k-1)g)}|\nu-1|^{-2(1+B+(k-1)g)}|\nu-z|^{-2+2kg}
  \;\mathfrak{J}_1^{(n-k)}(A+kg,B+kg|\nu)\,d^2\nu.
\end{multline}
In particular, for $k=n$ we have 
\begin{multline}\label{App-4point-recurent-n}
  \mathfrak{J}_1^{(n)}(A,B|z)\sim|z|^{2(1+A-g)}|z-1|^{2(1+B-g)}
  \int|\nu|^{-2(1+A+(n-1)g)}|\nu-1|^{-2(1+B+(n-1)g)}|\nu-z|^{-2+2ng}d^2\nu\sim\\\sim
  \int|\tau|^{2(A+(n-1)g)}|\tau-1|^{2(B+(n-1)g)}|\tau-z|^{-2ng}\,d^2\tau.
\end{multline}
In the last line in Eq \eqref{App-4point-recurent-n} we transformed integral over variable $\nu$ using identity \eqref{Bazeyaka}. Integral \eqref{App-4point-recurent-n} can be rewritten using definition of the kernel \eqref{K1} as
\begin{equation}\label{App-4point-recurent-n-kernel}
  \mathfrak{J}_1^{(n)}(A,B|z)\sim\int|\tau|^{2(A+(n-1)g)}|\tau-1|^{2(B+(n-1)g)}K_1^{-ng}(\tau|z)\,d^2\tau.
\end{equation}
Total factor in Eq \eqref{App-4point-recurent-n-kernel} can be obtained by setting $z=0$.
\paragraph*{Five-point function.}
Now we consider integral for the five-point function
\begin{equation}\label{5point-0step}
  \mathfrak{J}_2^{(n)}(A,B|z_1,z_2)=
  \int\prod_{k=1}^n 
  |t_k|^{2A}|t_k-1|^{2B}|t_k-z_1|^{-2g}|t_k-z_2|^{-2g}\mathcal{D}_n^{2g}(t)\,d^2\vec{t}_n.
\end{equation}
Substituting $\mathcal{D}_n^{2g}(t)$ into Eq \eqref{5point-0step} in the form \eqref{Bazeyaka-start} and performing integration over the variables $t_k$ using Eq \eqref{Bazeyaka} one can find the following relation between integrals $\mathfrak{J}_2^{(n)}$ and $\mathfrak{J}_2^{(n-1)}$
\begin{multline}\label{App-5point-recurent-1}
  \mathfrak{J}_2^{(n)}(A,B|z_1,z_2)\sim\prod_{k=1}^2|z_k|^{2(1+A-g)}|z_k-1|^{2(1+B-g)}|z_1-z_2|^{2-4g}
  \times\\\times\int\prod_{j=1}^2
  |\xi_j|^{-2(1+A)}|\xi_j-1|^{-2(1+B)}|\xi_j-z_1|^{-2+2g}|\xi_j-z_2|^{-2+2g}
  \;\mathfrak{J}_2^{(n-1)}(A+g,B+g|\xi_1,\xi_2)\mathcal{D}_2(\xi)\,d^2\vec{\xi}_2.
\end{multline}
If we apply relation \eqref{App-5point-recurent-1} twice we obtain that
\begin{multline}\label{App-5point-recurent-2}
  \mathfrak{J}_2^{(n)}(A,B|z_1,z_2)\sim\prod_{k=1}^2|z_k|^{2(1+A-g)}|z_k-1|^{2(1+B-g)}|z_1-z_2|^{2-4g}
  \times\\\times\int\prod_{j=1}^2
  |\nu_j|^{-2(1+A+g)}|\nu_j-1|^{-2(1+B+g)}
  \;\mathfrak{J}_2^{(n-2)}(A+2g,B+2g|\nu_1,\nu_2)\mathcal{D}_2(\nu)\,d^2\vec{\nu}_2\\
  \prod_{k=1}^2|\xi_k-\nu_1|^{-2+2g}|\xi_k-\nu_2|^{-2+2g}
  |\xi_k-z_1|^{-2+2g}|\xi_k-z_2|^{-2+2g}\mathcal{D}_2^{2-2g}(\xi)\,d^2\vec{\xi}_2.
\end{multline}
Integral in the last line of Eq \eqref{App-5point-recurent-2} can be reduced to the two-dimensional 
integral using the following identity\footnote{This simple integral identity reflects in particular the interesting fact that four-point correlation function of the fields $\Phi_{b/2}$ in GMM with central charge $c=1-6(b-b^{-1})^2$ (integral in the r.~h.~s.) coinsides up to a trivial factor with four-point correlation function of the fields $\Phi_{b'/2}$ in another GMM with central charge $c'$ corresponding to $b'^2=1-b^2$ (integral in the l.~h.~s.).}, which can be derived using Eqs \eqref{Bazeyaka} and \eqref{Star-Triangle}
\begin{multline}\label{Min_Identity}
  \int\prod_{k=1}^2|\xi_k-\nu_1|^{-2+2g}|\xi_k-\nu_2|^{-2+2g}|\xi_k-z_1|^{-2+2g}|\xi_k-z_2|^{-2+2g}
   \,\mathcal{D}_2^{2-2g}(\xi)\,d^2\xi_1d^2\xi_2=2\pi\frac{\gamma^5(g)\gamma(2-3g)}{\gamma(2g)}
   \\\times
   |\nu_1-\nu_2|^{4g-2}|\nu_1-z_1|^{4g-2}|\nu_2-z_1|^{4g-2}\int
   |y-z_1|^{-2g}|y-z_2|^{6g-4}|y-\nu_1|^{-2g}|y-\nu_2|^{-2g}\,d^2y.
\end{multline}
As a result of application of the identity \eqref{Min_Identity} one obtains
\begin{multline}\label{App-5point-recurent-2-final}
  \mathfrak{J}_2^{(n)}(A,B|z_1,z_2)\sim\prod_{k=1}^2|z_k|^{2(1+A-g)}|z_k-1|^{2(1+B-g)}|z_1-z_2|^{2-4g}
  \times\\\times\int\prod_{j=1}^2
  |\nu_j|^{-2(1+A+g)}|\nu_j-1|^{-2(1+B+g)}|\nu_j-z_1|^{4g-2}
  \;\mathfrak{J}_2^{(n-2)}(A+2g,B+2g|\nu_1,\nu_2)\mathcal{D}_2^{2g}(\nu)\,d^2\vec{\nu}_2\\
   |y-z_1|^{-2g}|y-z_2|^{6g-4}|y-\nu_1|^{-2g}|y-\nu_2|^{-2g}\,d^2y.
\end{multline}
Now Eq \eqref{App-5point-recurent-2-final} is stabilized. It means, that after applying relation \eqref{App-5point-recurent-1} the auxiliary integrations can be performed and remaining integral will have the similar form and only parameters in the integrand will change:
\begin{multline}\label{App-5point-recurent-3}
  \mathfrak{J}_2^{(n)}(A,B|z_1,z_2)\sim\prod_{k=1}^2|z_k|^{2(1+A-g)}|z_k-1|^{2(1+B-g)}|z_1-z_2|^{2-4g}
  \times\\\times\int\prod_{j=1}^2
  |\nu_j|^{-2(1+A+2g)}|\nu_j-1|^{-2(1+B+2g)}|\nu_j-z_1|^{6g-2}
  \;\mathfrak{J}_2^{(n-3)}(A+3g,B+3g|\nu_1,\nu_2)\mathcal{D}_2^{2g}(\nu)\,d^2\vec{\nu}_2\\
   |y-z_1|^{-4g}|y-z_2|^{8g-4}|y-\nu_1|^{-2g}|y-\nu_2|^{-2g}\,d^2y.
\end{multline}
If we repeat this procedure $n$ times we obtain that
\begin{multline}\label{App-5point-recurent-n}
  \mathfrak{J}_2^{(n)}(A,B|z_1,z_2)\sim\prod_{k=1}^2|z_k|^{2(1+A-g)}|z_k-1|^{2(1+B-g)}|z_1-z_2|^{2-4g}
  \times\\\times\int\prod_{j=1}^2
  |\nu_j|^{-2(1+A+(n-1)g)}|\nu_j-1|^{-2(1+B+(n-1)g)}|\nu_j-z_1|^{2ng-2}
  \mathcal{D}_2^{2g}(\nu)\,d^2\vec{\nu}_2\\
   |y-z_1|^{-2(n-1)g}|y-z_2|^{2(n+1)g-4}|y-\nu_1|^{-2g}|y-\nu_2|^{-2g}\,d^2y.
\end{multline}
Using identities \eqref{Bazeyaka} and \eqref{App-int-4point} integral in the r.~h.~s. of Eq \eqref{App-5point-recurent-n} can be transformed to
\begin{multline}\label{App-5point-recurent-n-final}
  \mathfrak{J}_2^{(n)}(A,B|z_1,z_2)\sim
  \int\prod_{j=1}^2
  |\nu_j|^{2(A+(n-2)g)}|\nu_j-1|^{2(B+(n-2)g)}|\nu_j-z_1|^{-2(n-1)g}
  \mathcal{D}_2(\nu)\,d^2\vec{\nu}_2\\
   |y-z_1|^{2(n-2)g}|y-z_2|^{-2ng}|y-\nu_1|^{-2+2g}|y-\nu_2|^{-2+2g}\,d^2y.
\end{multline}
Eq \eqref{App-5point-recurent-n-final} can be expressed in terms of integral with kernel \eqref{K2}
\begin{equation}\label{App-5point-recurent-n-kernel}
  \mathfrak{J}_2^{(n)}(A,B|z_1,z_2)\sim
  \int\prod_{j=1}^2
  |\nu_j|^{2(A+(n-2)g)}|\nu_j-1|^{2(B+(n-2)g)}
  K_2^{-(n-1)g}(\nu_1,\nu_2|z_1,z_2)
  \mathcal{D}_2^{2g}(\nu)\,d^2\vec{\nu}_2.
\end{equation}
\paragraph*{$\mathbf{m+3}$-point function.}
Simplification of the integral \eqref{App-mpoint} for general $m$ can be done in the same way.
Integral relations \eqref{App-4point-recurent-1} and \eqref{App-5point-recurent-1} can be generalized:
\begin{multline}\label{App-mpoint-recurent-1}
  \mathfrak{J}_m^{(n)}(A,B|z_1,..,z_m)\sim\prod_{k=1}^m|z_k|^{2(1+A-g)}|z_k-1|^{2(1+B-g)}
  \prod_{i<j}|z_i-z_j|^{2-4g}
  \times\\\times\int\prod_{j=1}^m
  |\xi_j|^{-2(1+A)}|\xi_j-1|^{-2(1+B)}\prod_{k=1}^m|\xi_j-z_k|^{-2+2g}
  \;\mathfrak{J}_m^{(n-1)}(A+g,B+g|\xi_1,..,\xi_m)\mathcal{D}_m(\xi)\,d^2\vec{\xi}_m.
\end{multline} 
Applying this relation $m$ times one can show, that integral stabilizes\footnote{To do that one should use identity, which generalizes identity \eqref{Min_Identity} for bigger number of "external" points $z_i$ and $\nu_j$.}. It means, that after the next applications of the relation \eqref{App-mpoint-recurent-1} it will have the same form (with shifted parameters $A$ and $B$), but with smaller number of integrations.  A final formula will look similar to \eqref{App-4point-recurent-n-kernel} and \eqref{App-5point-recurent-n-kernel}
\begin{multline}\label{App-mpoint-recurent-n-kernel}
  \mathfrak{J}_m^{(n)}(A,B|z_1,..,z_m)\sim\\\sim
  \int\prod_{j=1}^m
  |\nu_j|^{2(A+(n-m)g)}|\nu_j-1|^{2(B+(n-m)g)}
  K_m^{-(n-m+1)g}(\nu_1,..,\nu_m|z_1,..,z_m)
  \mathcal{D}_m^{2g}(\nu)\,d^2\vec{\nu}_m,
\end{multline} 
where $K_m^{\Delta}(\nu_1,..,\nu_m|z_1,..,z_m)$ is given by Eq \eqref{Kernel}.
\section{Useful integral identities}
In this appendix we give without a proof useful integral relations, which permit to reduce essentially the number of integrations in correlation function \eqref{n-point} and help to simplify calculation of correlation functions.
\subsection*{Generalization of relation \eqref{Bazeyaka}.}
Instead of relation \eqref{Bazeyaka} sometimes it is convenient to use relation 
\begin{multline}\label{Star-Triangle}
   \int\mathcal{D}_n(y)\prod_{i=1}^n\prod_{j=1}^{n+m+2}|y_i-t_j|^{2p_j}
   \,d^2\vec{y}_n=\prod_{j=1}^{n+m+2}\gamma(1+p_j)
   \prod_{i<j}|t_i-t_j|^{2+2p_i+2p_j}\times\\\times
   \int\mathcal{D}_m(u)\prod_{i=1}^m\prod_{j=1}^{n+m+2}|u_i-t_j|^{-2-2p_j}
   \,d^2\vec{u}_m,
\end{multline}
where $\sum_kp_k=-n-1$. 

This relation generalizes well known star-triangle relation, which corresponds to the case $m=0$ and $n=1$. The equation for asymptotic \eqref{y_k=y} of the kernel \eqref{Kernel} can be easily derived by successive application of this relation. The same relation helps to establish the fusion procedure in the correlation function \eqref{n-point}.

The following identity can be used to prove the symmetry properties of the kernel \eqref{K2}
\begin{multline}\label{App-int-4point}
  \int|t-\xi_1|^{2p_1}|t-\xi_2|^{2p_2}|t-\xi_3|^{2p_3}|t-\xi_4|^{2p_4}d^2t=
  \prod_{k=1}^4\gamma(1+p_k)
  |\xi_1-\xi_2|^{2+2p_1+2p_2}|\xi_3-\xi_4|^{2+2p_3+2p_4}\times\\\times
  \int|s-\xi_1|^{-2-2p_2}|s-\xi_2|^{-2-2p_1}|s-\xi_3|^{-2-2p_4}|s-\xi_4|^{-2-2p_3}d^2s,
\end{multline}
where $\sum p_k=-2$. 

This identity  can be applied for any two pairs of points $\xi_j$ (in Eq \eqref{App-int-4point} they are chosen as $\xi_1,\xi_2$ and $\xi_3,\xi_4$). Applying it three times, we obtain relation \eqref{Star-Triangle} for $m=1$ and $n=1$.
\subsection*{Factorization property of  the integral \eqref{n-point}.}\label{Appendix-J}
There is useful property of the integral \eqref{n-point}
\begin{multline}\label{Kernel-razbienie}
  \int\prod_{k=1}^n\vert t_k\vert^{2A}\vert t_k-1\vert^{2B}K_n^{\Delta}(t_1,..,t_n|z_1,..,z_n)
  \mathcal{D}_n^{2g}(t)d^2\vec{t}_n=
  \Lambda_k
  \prod_{j=n-k+1}^n\hspace*{-10pt}|z_j|^{2(1+A+\Delta)}|z_j-1|^{2(1+B+\Delta)}\\
  \begin{aligned}[c]
   \times&\int\prod_{j=1}^{n-k}\vert t_j\vert^{2(A+kg)}\vert t_j-1\vert^{2(B+kg)}
    K_{n-k}^{\Delta-kg}(t_1,..,t_{n-k}|z_1,..,z_{n-k})
  \mathcal{D}_{n-k}^{2g}(t)\,d^2\vec{t}_{n-k}\prod_{ij}|t_i-y_j|^{-2g}\\
   \prod_{j=1}^k&\vert y_j\vert^{-2(1+A+(k-1)g)}\vert y_j-1\vert^{-2(1+A+(k-1)g)}
   K_k^{-1-\Delta+(k-1)g}(y_1,..,y_k|z_{n-k+1},..,z_n)
  \mathcal{D}_k^{2g}(y)\,d^2\vec{y}_k
  \end{aligned}
\end{multline}
with
\begin{equation*}
 \Lambda_k(A,B,\Delta)=\prod_{j=0}^{k-1}\frac{\gamma((n-j)g)}{\gamma((j+1)g)}
  \frac{\gamma(1+A+jg)\gamma(1+B+jg)\gamma(1+\Delta-jg)}{\gamma(2+A+B+\Delta+(m-1-j)g)}.
\end{equation*}
This property allows to reduce essentially the number of integrations in the correlation function
\begin{equation}\label{Number_of_Int_in_corr_func}
  \langle V_{-\frac{m_1b}{2}}(z_1)\dots V_{-\frac{m_kb}{2}}(z_k)
   V_{\alpha_1}(0)V_{\alpha_2}(1)V_{\alpha_3}(\infty)\rangle.
\end{equation}
Let us suppose, that $m_1\geq m_2\geq\dots\geq m_k$, then using factorization property \eqref{Kernel-razbienie} and Eq \eqref{Star-Triangle} one can show that number of integrations $N$ in correlation function \eqref{Number_of_Int_in_corr_func} can be reduced up to:
\begin{equation}\label{N-number}
   \frac{N}{2}=\sum_{j=1}^{[k/2]}j(m_{2j}+m_{2j-1})+\frac{m_k(k+1)}{4}(1-(-1)^k).
\end{equation}
In particular for $m_1=m_2=\dots=m_k=1$ the number of integrations \eqref{N-number} in correlation function \eqref{Number_of_Int_in_corr_func} is equal to
\begin{equation}
  N=\frac{k(k+2)}{2}+\frac{1}{4}(1-(-1)^k).
\end{equation} 
We note, that the number of integrations in Eq \eqref{n-point} for this correlation function is equal to $k(k+1)$.
\subsection*{Analytical properties of the integral \eqref{4point}.}\label{4point-integrals}
It is convenient to consider more general integral (integral \eqref{4point} will correspond to: $\xi_1=0$, $\xi_2=1$, $\xi_3=z$, $p_1=A$, $p_2=B$, $p_3=C$ and $g=-b^2$): 
\begin{equation}
\mathbf{J}_m(p_1,p_2,p_3|\xi_1,\xi_2,\xi_3)=
  \int\prod_{k=1}^m
  \vert t_k-\xi_1\vert^{2p_1}\vert t_k-\xi_2\vert^{2p_2}
  \vert t_k-\xi_3\vert^{2p_3}\mathcal{D}^{2g}_m(t)\,d^2\vec{t}_m.
\end{equation}
This integral satisfies the following remarkable relations,  which can be used for the analytical continuation of $\mathbf{J}_m(p_1,p_2,p_3|\xi_1,\xi_2,\xi_3)$ as a function of parameters $p_1$, $p_2$ and $p_3$. Namely, for any pair of points $\xi_j$ (for example $\xi_1$ and $\xi_2$) one has
\begin{multline}\label{App-4point-relation}
  \mathbf{J}_m(p_1,p_2,p_3|\xi_1,\xi_2,\xi_3)=G_m(p_1,p_2,p_3)\;
  |\xi_1-\xi_2|^{2m(1+p_1+p_2+(m-1)g)}\times\\\times
  \mathbf{J}_m(-1-p_2-(m-1)g,-1-p_1-(m-1)g,-1-p_4-(m-1)g|\xi_1,\xi_2,\xi_3)
\end{multline}
with
\begin{equation*}
 G_m(p_1,p_2,p_3)=\prod_{j=0}^{m-1}
 \gamma(1+p_1+jg)\gamma(1+p_2+jg)\gamma(1+p_3+jg)\gamma(1+p_4+jg)
\end{equation*}
where we have introduced the notation: $p_4=-2-p_1-p_2-p_3-2(m-1)g$. 

Applying this relation three times one obtains:
\begin{multline}\label{App-4point-relation-2}
  \mathbf{J}_m(p_1,p_2,p_3|\xi_1,\xi_2,\xi_3)=G_m(p_1,p_2,p_3)\;
  \prod_{i<j}|\xi_i-\xi_j|^{2m(1+p_i+p_j+(m-1)g)}\times\\\times
  \mathbf{J}_m(-1-p_1-(m-1)g,-1-p_2-(m-1)g,-1-p_3-(m-1)g|\xi_1,\xi_2,\xi_3)
\end{multline}
Considering the analytical properties of left and right hand sides of Eqs \eqref{App-4point-relation} and
\eqref{App-4point-relation-2} one can conclude, that function
\begin{equation}
  \widetilde{\mathbf{J}}_m(p_1,p_2,p_3|\xi_1,\xi_2,\xi_3)=\prod_{k=1}^4\prod_{j=0}^{m-1}
  \Gamma^{-1}(1+p_k+jg)\,\mathbf{J}_m(p_1,p_2,p_3|\xi_1,\xi_2,\xi_3)
\end{equation}
is entire function of the parameters $p_1$, $p_2$ and $p_3$ for all finite and not coinsiding points $\xi_k$. Similar but more tedious relations can be derived  also for the integral \eqref{Jmm}.

\end{document}